# Applying ANN, ANFIS, and LSSVM Models for Estimation of Acid Solvent Solubility in Supercritical CO2

*Amin Bemani[1], Alireza Baghban[2], Shahaboddin Shamshirband[3,4], Amir Mosavi[5,6], Peter Csiba[7], Annamária R. Várkonyi-Kóczy[5,7]*

[1] *Petroleum Engineering Department, Petroleum University of Technology, Ahwaz, Iran; aminbemani90@yahoo.com*
[2] *Chemical Engineering Department, Amirkabir University of Technology, Mahshahr Campus, Mahshahr, Iran; baghban1369@gmail.com*
[3] *Department for Management of Science and Technology Development, Ton Duc Thang University, Ho Chi Minh City, Viet Nam*
[4] *Faculty of Information Technology, Ton Duc Thang University, Ho Chi Minh City, Viet Nam*
[5] *Department of Automation, Óbuda University, Budapest, Hungary*
[6] *School of the Built Environment, Oxford Brookes University, Oxford OX3 0BP, UK*
 *amir.mosavi@kvk.uni-obuda.hu*
[7] *Department of Mathematics and Informatics, J. Selye University, Komarno 94501, Slovakia csibap@ujs.sk and koczya@ujs.sk*

**Abstract**

In the present work, a novel and the robust computational investigation is carried out to estimate solubility of different acids in supercritical carbon dioxide. Four different algorithms such as radial basis function artificial neural network, Multi-layer Perceptron (MLP) artificial neural network (ANN), Least squares support vector machine (LSSVM) and adaptive neuro-fuzzy inference system (ANFIS) are developed to predict the solubility of different acids in carbon dioxide based on the temperature, pressure, hydrogen number, carbon number, molecular weight, and acid dissociation constant of acid. In the purpose of best evaluation of proposed models, different graphical and statistical analyses and also a novel sensitivity analysis are carried out. The present study proposed the great manners for best acid solubility estimation in supercritical carbon dioxide, which can be helpful for engineers and chemists to predict operational conditions in industries.

## 1 Introduction

In the recent years, supercritical fluid has become one of the interests of chemical engineers and chemists as a novel and extensive applicable technology. The synthesis and generating of nanomaterials and extraction process of different materials are the popular applications of supercritical fluids (Inomata, Honma et al. 1999, Stassi, Bettini et al. 2000, Ohde, Hunt et al. 2001, Celso, Triolo et al. 2002, Üzer, Akman et al. 2006, Munshi and Bhaduri 2009, Nahar and Sarker 2012, Zhang, Heinonen et al. 2014, Knez, Cör et al. 2017, Zhao, Zhang et al. 2017, Belghait, Si-Moussa et al. 2018, Daryasafar, Daryasafar et al. 2018, Gao, Abdi-khanghah et al. 2018). One of the supercritical fluids which have wide applications in the extraction of various metals from solid and liquid phases is carbon dioxide (Erkey 2000, Sunarso and Ismadji 2009, Lin, Liu et al. 2014). Due to non-flammability, nontoxicity, low cost, and critical points (304.2 K and 7.38 MPa) of carbon dioxide, the supercritical carbon dioxide becomes one of the interesting and applicable supercritical fluids in industries (Ghaziaskar and Nikravesh 2003, Bovard, Abdi et al. 2017). The viscosity and density of supercritical carbon dioxide are known as two important transport properties of the fluids which are affected by pressure and temperature. Another dominant thermos physical property of supercritical carbon dioxide is solubility of different materials in supercritical carbon dioxide which is a function of various factors such as polarity, molecular weight, pressure, temperature, and vapor pressure (Huang, Chiew et al. 2005, Ghaziaskar and Kaboudvand 2008). One types of the materials which have a solubility in supercritical carbon dioxide are acids, the nanofluoropentanoic acid which is known as one type of perfluorocarboxylic acids, has extensive applications in the production of paints additives, polymers, foams, and stain repellents but because of their high ability instability they are harmful to environment (Richter and Dibble 1983, Moody and Field 1999, Hintzer, Löhr et al. 2004, Fei and Olsen 2011, Hubbard, Guo et al. 2012, Dartiguelongue, Leybros et al. 2016, Hintzer, Juergens et al. 2016). Adrien Dartiguelongue and coworkers studied solubility of perfluoropentanoic acid in supercritical carbon dioxide in the wide range of temperature and pressure and also proposed some density based models to predict solubility in terms of density of supercritical fluids (Dartiguelongue, Leybros et al. 2016). Gurdial et al. constructed dynamic setup to study solubility of o-, m- and p-hydroxybenzoic acid in the supercritical carbon dioxide in the wide range pressure of 80-205 mbar and temperature range of 308.15-328.15 K and correlated the measured solubility as a function of density (Gurdial and Foster 1991). Kumoro measured the solubility of 2R,3β-dihydroxyurs-12-en-28-oic acid which is called Corosolic acid dynamically in a different range of pressure 8 to 30 MPa and five different temperatures of 308.15, 313.15, 323.15, and 333.15 K. Kumoro used various density based models to correlate the experimental data (Kumoro 2011).

Sahihi et al. measured the solubility of Maleic acid in supercritical carbon dioxide by utilization of static experimental setup. The measured data belongs to Maleic acid in pressure range of 7 to 300 bar and temperature of 348.15 K (Sahihi, Ghaziaskar et al. 2010). Ghaziaskar and coworkers used a continuous flow set up to study solubility of tracetin, diacentin and acetic acid in supercritical carbon dioxide in the pressure range of 70 to 180 bar and various temperature of 313, 333 and 348 K and they also compared the obtained solubilities for different acids (Ghaziaskar, Afsari et al. 2017). Helena Sovova adjusted the Adachi-Lu equation based on the solubility of Ribes nigrum (blackcurrant) and Vitis vinifera (grape-vine) in supercritical carbon dioxide. They concluded the Adachi-Lu equation has enough accuracy in forecasting solubility of triglycerides in carbon dioxide (Sovova, Zarevucka et al. 2001).

The issue of prediction of various acids solubility in supercritical carbon dioxide and phase equilibrium investigation of supercritical carbon dioxide and different materials are the important topics in chemical engineering research. According to the hardships of experimental studies such as special tools and procedure which are needed, in the present work, the mathematical investigation is considered as a great solution for these problems (Anitescu, Atroshchenko et al. 2019, Guo, Zhuang et al. 2019, Rabczuk, Ren et al. 2019, Zarei, Razavi et al. 2019). In this paper four different algorithms, Radial basis function artificial neural network (RBF-ANN), Multi-layer Perceptron artificial neural network (MLP-ANN), Least squares support vector machine (LSSVM) and Adaptive neuro-fuzzy inference system (ANFIS) are developed to predict the solubility of different types of acid in supercritical carbon dioxide based on the various parameters such as structure of acid, pressure and temperature.

## 2 Methodology

### 2.1 Experimental Data Gathering

The dominant purpose of present paper is development of accurate and simple models to forecast solubility of different acids in supercritical carbon dioxide. Due to this, the required actual data for training and testing phases of models were assembled from the reliable source existed in literature (Gurdial and Foster 1991, Sovova, Zarevucka et al. 2001, Sparks, Hernandez et al. 2007, Tian, Jin et al. 2007, Sparks, Estévez et al. 2008, Kumoro 2011, Dartiguelongue, Leybros et al. 2016). This collection of data contains the 188 acid solubility data points in terms of pressure, temperature, acid dissociation constant, molecular weight, number of carbon and hydrogen of acid. The details of data collection are reported in Tab. S1 and Tab. S2. This details include acid name, acid dissociation constant, pressure and temperature ranges and number of utilized data points for each acid. Also, for clarification of this experimental dataset, the structure, linear formula and molecular weight of utilized acids are presented in **Tab. S3**. These acids include Perfluoropentanoic acid, o-Hydroxybenzoic Acid, Corosolic Acid, Maleic Acid, Ferulic Acid, Azelaic Acid, p-aminobanzoic acid and Nonanioc acid.

### *2.2 Artificial neural networks*

Artificial neural networks have amazing similarities to the performance and structure of neuron units in the brain system (Smith 1993, Baş and Boyacı 2007). These computational blocks construct different types of layer such as input, output and hidden layers. In the layers, there are transfer functions or activation function which organize the process of training in the algorithm. Each neuron has specific weight and bias values which control the optimization process. Artificial neural network has ability of tracing a nonlinear form relationship between input and output parameters. Due to this ability, artificial neural networks have widespread application in different industries and sciences.

Artificial neural networks can be classified in different forms such as a recurrent neural network (RNN), radial basis function and multilayer perceptron (Movagharnejad, Mehdizadeh et al. 2011, Abdi-Khanghah, Bemani et al. 2018, Zamen, Baghban et al. 2019). In the present work, the MLP and RBF network are utilized.

### *2.3 Least squares support vector machine*

Vapnik organized support vector machine based on statistical learning theory (Vapnik 1998). This computational intelligence can be used for regression and classification purposes. However, there are many advantages to this method but there is a hardship in its computational procedure because of quadratic programming. The least squares SVM (LSSVM) is proposed as a novel type of SVM to solve this problem. This novel approach organized linear equations for computation and optimization (Cortes and Vapnik 1995, Suykens and Vandewalle 1999, Suykens, Vandewalle et al. 2001, Zamen, Baghban et al. 2019).

By considering a dataset of $(x_i, y_i)_n$, the LSSVM regression prediction is utilized to estimate a function, where $x_i$ and $y_i$ are known as input and target parameters and n represent the number of data which utilized in training phase(Wang, Zhang et al. 2005). The linear regression is formulated such as following:

$$y = \omega^T \varphi(x) + b \quad \quad \text{Eq. (1)}$$

Where $\varphi(x)$ denotes a nonlinear function that has different forms such as polynomial, linear, sigmoid and radial basis functions. Also, $\omega$ and b denote the weights and determined constant coefficient in training process. A new optimization problem can be defined based on LSSVM approach (Baghban, Bahadori et al. 2016, Baghban, Namvarrechi et al. 2016, Ahmadi, Baghban et al. 2019):

$$\min_{\omega,b,e} J(\omega, e) = \frac{1}{2}\omega^T\omega + \frac{1}{2}\gamma \sum_{k=1}^{N} e_k^2 \qquad \text{Eq. (2)}$$

Which is related to the below constraints:

$$y_k = \omega^T \varphi(x_k) + b + e_k \qquad k=1,2,\ldots,N \qquad \text{Eq. (3)}$$

The Lagrangian equation is constructed to solve the optimization problem:

$$L(\omega, b, e, \alpha) = J(\omega, e) - \sum_{k=1}^{N} \alpha_k \{\omega^T \varphi(x_k) + b + e_k - y_k\} \qquad \text{Eq. (4)}$$

Where ϒ and $e_k$ are known as regularization parameter and regression error. The $\alpha_k$ represent the support value. To solve the above problem, the above equation is differentiated with respect to the different parameters:

$$\frac{\partial L(\omega,b,e,\alpha)}{\partial \omega} = 0 \rightarrow \omega = \sum_{k=1}^{N} \alpha_k \varphi(x_k) \qquad \text{Eq. (5)}$$

$$\frac{\partial L(\omega,b,e,\alpha)}{\partial b} = 0 \rightarrow \sum_{k=1}^{N} \alpha_k = 0 \qquad \text{Eq. (6)}$$

$$\frac{\partial L(\omega,b,e,\alpha)}{\partial e_k} = 0 \rightarrow \alpha_k = \gamma e_k, \quad k=1,2,\ldots,N \qquad \text{Eq. (7)}$$

$$\frac{\partial L(\omega,b,e,\alpha)}{\partial \alpha_k} = 0 \rightarrow y_k = \omega^T \varphi(x_k) + b + e_k \quad k=1,2,\ldots,N \qquad \text{Eq. (8)}$$

Karush–Kuhn–Trucker matrix can be obtained by elimination of $\omega$ and $e$ (Cortes and Vapnik 1995, Baylar, Hanbay et al. 2009, Mehdizadeh and Movagharnejad 2011):

$$\begin{bmatrix} 0 & 1_v^T \\ 1_v & \Omega + \gamma^{-1}I \end{bmatrix} \begin{bmatrix} b \\ \alpha \end{bmatrix} = \begin{bmatrix} 0 \\ y \end{bmatrix} \qquad \text{Eq. (9)}$$

where $y = [y_1 \ldots y_N]^T$, $\alpha = [\alpha_1 \ldots \alpha_N]^T$, $1_N = [1 \ldots 1]^T$, and I represents the identity matrix. $\Omega_{kl}$ is $\varphi(x_k)^T \varphi(x_l) = K(x_k, x_l)$. K($x_k, x_l$) is known as kernel function which can be in different forms of linear, polynomial and radial basis function forms(Gunn 1998). The estimating function form of LSSVM algorithm can be expressed as following formulation(Muller, Mika et al. 2001, Rostami, Baghban et al. 2019):

$$y(x) = \sum_{k=1}^{N} \alpha_k K(x, x_k) + b \qquad \text{Eq. (10)}$$

### *2.4 Adaptive neuro-fuzzy inference system (ANFIS)*

Adaptive neuro-fuzzy inference system which is called ANFIS algorithm, in brief, has five different layers. The aforementioned approach was developed by Jang and Sun(Jang, Sun et al. 1997). The hybrid learning approach and back propagation are known as fundamentals of training of conventional ANFIS algorithm. The ANFIS algorithm was born base on fuzzy logic and neural network advantages and also the different evolutionary methods such as Imperialist Competitive Algorithm (ICA), Particle Swarm Optimization (PSO) and Genetic algorithm (GA) can be used to reach the optimal structure of ANFIS algorithm(Afshar, Gholami et al. 2014, Khosravi, Nunes et al. 2018, Razavi, Sabaghmoghadam et al. 2019). The ANFIS structure is demonstrated in **Fig. 1**. As shown there are two input variables and one output.

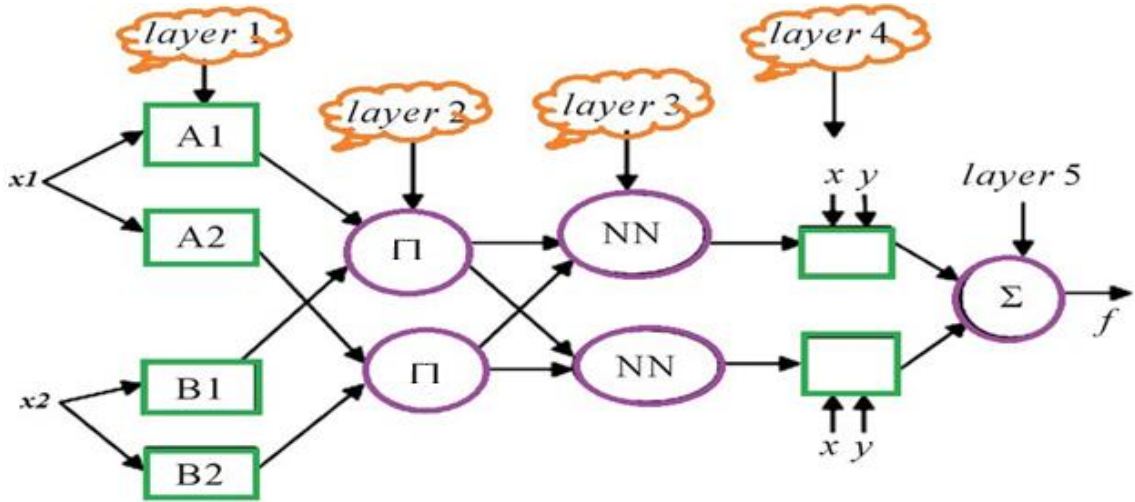

**Figure 1:** Typical construction of ANFIS approach

In the first layer, the linguistic terms are built based on input data. The Gaussian membership function is applied to organize these linguistic terms. The Gaussian function can be shown as following formulation (Ahangari, Moeinossadat et al. 2015, Bahadori, Baghban et al. 2016):

$$O_i^1 = \beta(X) = exp^{(-\frac{1}{2}\frac{(X-Z)^2}{\sigma^2})} \qquad \text{Eq. (11)}$$

Where Z and σ denote the Gaussian parameters.

The next layer, shown as Π multiplies the incoming signals and contains the weighted terms which are related to rules:

$$O_i^2 = W_i = \beta_{Ai}(X).\beta_{Bi}(X) \qquad \text{Eq. (12)}$$

The third layer the shown as NN, it averages of determined weights are evaluated such as the following formulation:

$$O_i^3 = \frac{W_i}{\sum W_i} \qquad \text{Eq. (13)}$$

Then in the next layer, the average weight values are multiplied to the related function such as below:

$$O_i^4 = \overline{W_i} f_i = \overline{W_i}(m_i X_1 + n_i X_2 + r_i) \qquad \text{Eq. (14)}$$

Where, m, n, and r represent the resulting indexes.

At last, the fifth layer consists of the summation of previous layer outputs:

$$O_i^5 = Y = \sum_i \overline{W_i} f_i = \overline{W_1} f_1 + \overline{W_2} f_2 = \frac{\sum W_i f_i}{\sum W_i} \qquad \text{Eq. (15)}$$

### *2.5 Particle swarm optimization (PSO)*

The combination of random probability distribution approach and generation of the population constructed the particle swarm optimization algorithm. Eberhart et al. introduced the PSO algorithm that comes from the social behavior of birds and developed it to solve nonlinear function optimization problems (Kennedy 2010). This strategy has special similarities with other optimization approach such as genetic algorithm which is constructed base on random solution population. Each particle can be known as a probable solution of problem. A random population of particle created in search space to relate in optimum system. $P_{best}$ is known as the best solution which can obtained from this strategy for a particle. Also $g_{best}$ represents the global best solution determined by swarm. The particle move in the space by time iterations and the next iteration velocity is determined by using $g_{best}$, $P_{best}$ and current velocity (Eberhart and Kennedy 1995). The P'th particle can be determined as follow:

$$X_{pd}^{iter+1} = X_{pd}^{iter} + V_{pd}^{iter+1} \qquad \text{Eq. (16)}$$

The particle velocity is updated by the following expression:

$$v_{id}(t+1) = w v_{id}(t) + c_1 r_1 \left( p_{best,id}(t) - X_{iid}(t) \right) + c_2 r_2 \left( g_{best,d}(t) - X_{id}(t) \right) \qquad \text{Eq. (17)}$$

*w, c,* and *r* are inertia weight, learning rate and random number respectively (Haratipour, Baghban et al. 2017).

### 3 Results and discussion

In the present study, the determined structure of MLP-ANN algorithm utilizes log-sigmoid and linear activation functions the hidden and output layers respectively. By utilization of trial and error, the optimum number of neurons in hidden layers is determined as 7 to reach the best structure of MLP-ANN algorithm. The performance of Levenberg Marquardt training of MLP-ANN algorithm based on the mean square error is shown in **Fig. 2**.

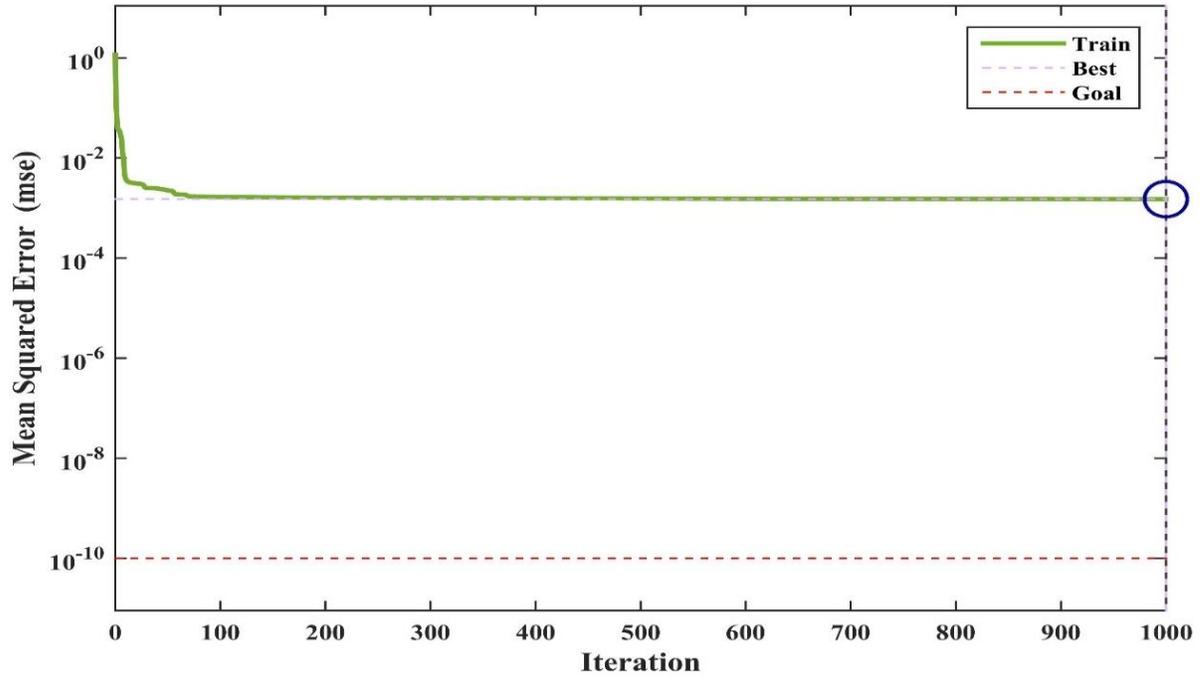

**Figure 2:** Trained MLP-ANN model by Levenberg Marquardt algorithm

In the RBF-ANN algorithm, the radial basis function (RBF) is utilized for hidden layers. According to information in the literature, the hidden layer neurons for RBF-ANN can be supposed one-tenth of training data points. The training process of RBF-ANN algorithm base on MSE has been reported in Fig. 3.

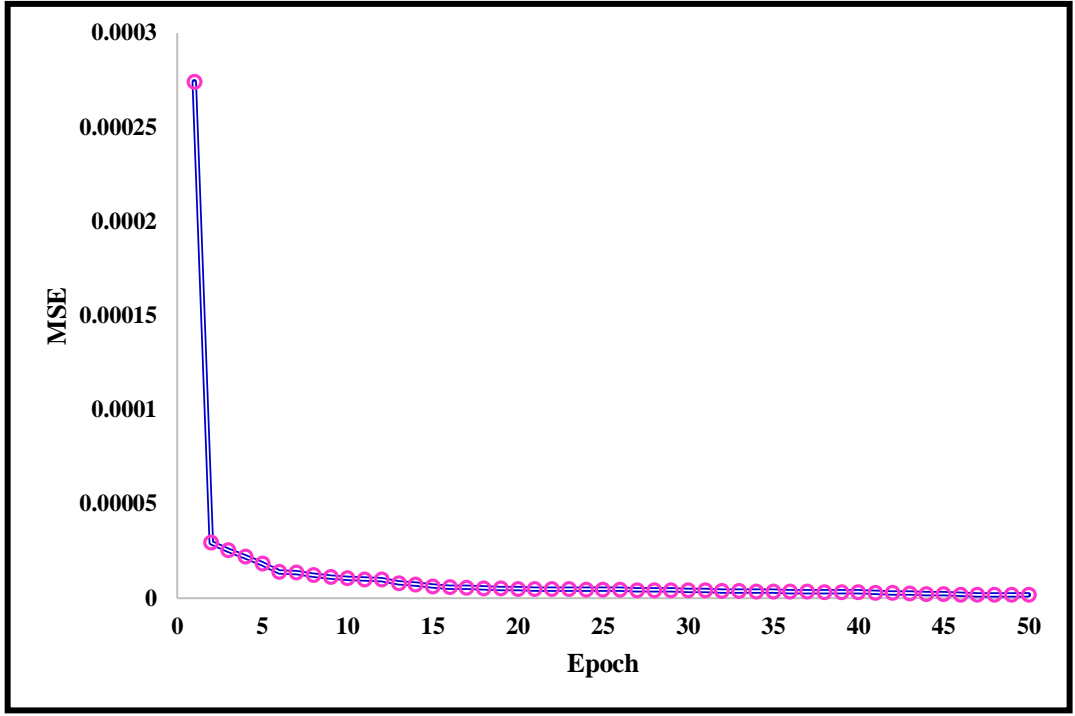

**Figure 3:** Trained RBF-ANN approach by Levenberg Marquardt algorithm

In this work, particle swarm optimization approach is applied to train the best structure of ANFIS algorithm. **Fig. 4** demonstrates the gained root mean squared error (RMSE) of estimated and experimental acid solubility values in training step.

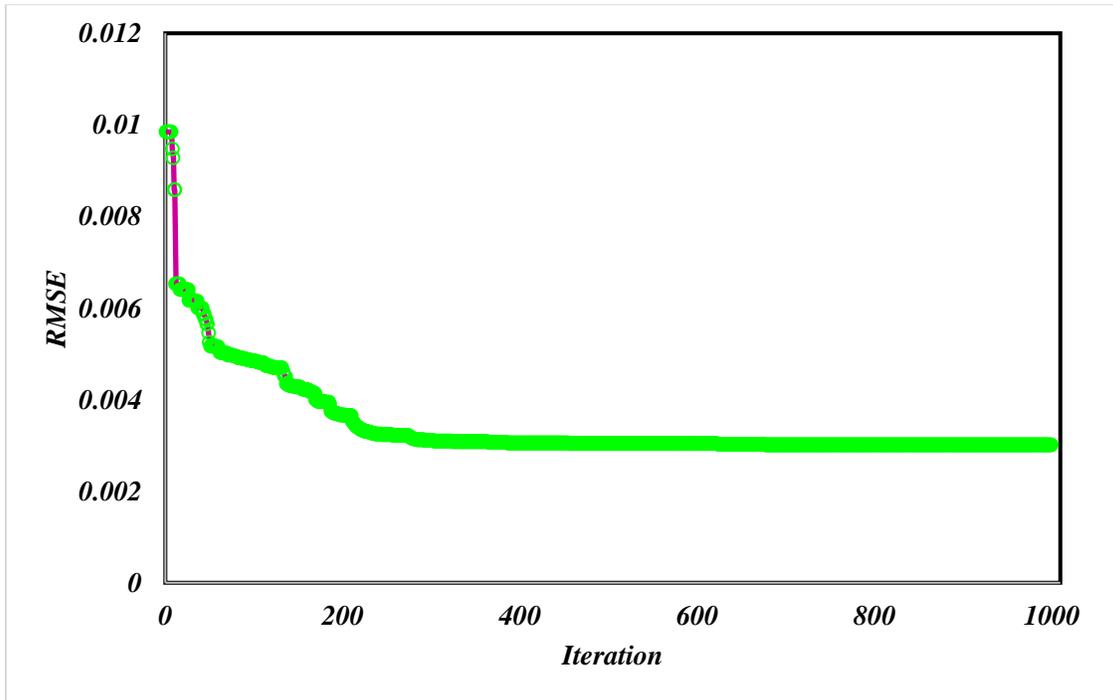

**Figure 4:** Performance of trained ANFIS model

The optimum structure of ANFIS can be recognized by the RMSE value of 0.003 after 1000 of iteration steps. Trained membership functions of proposed ANFIS model are also shown in **Fig. 5** for each cluster.

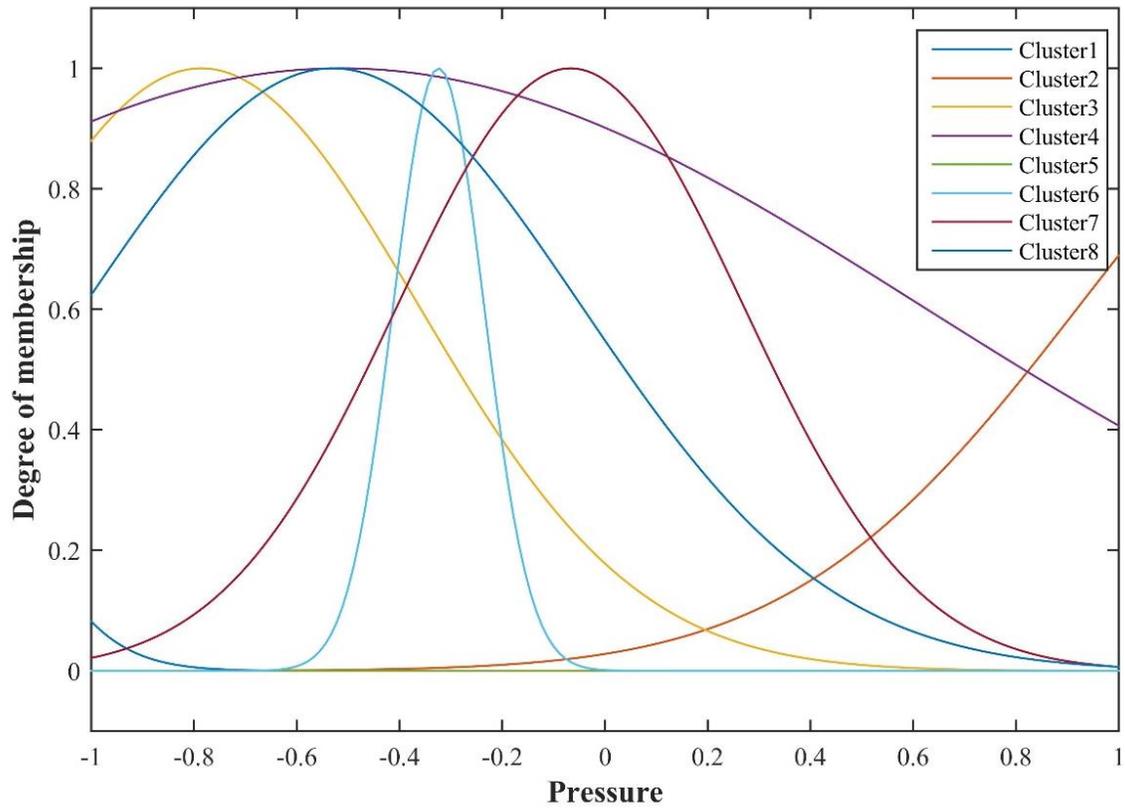

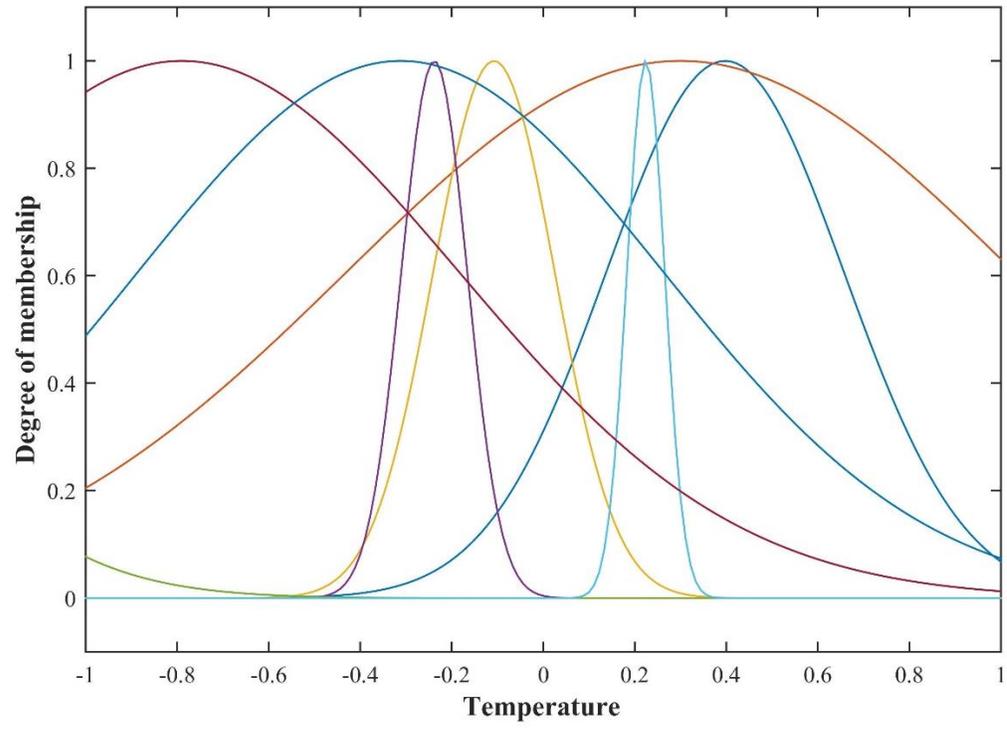

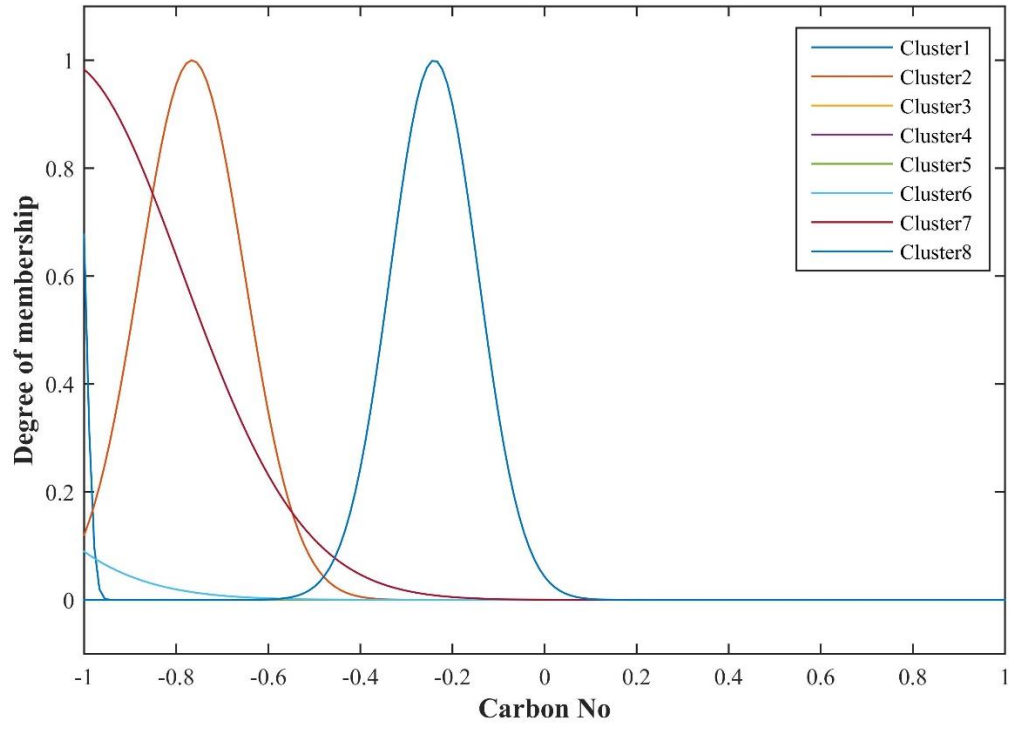

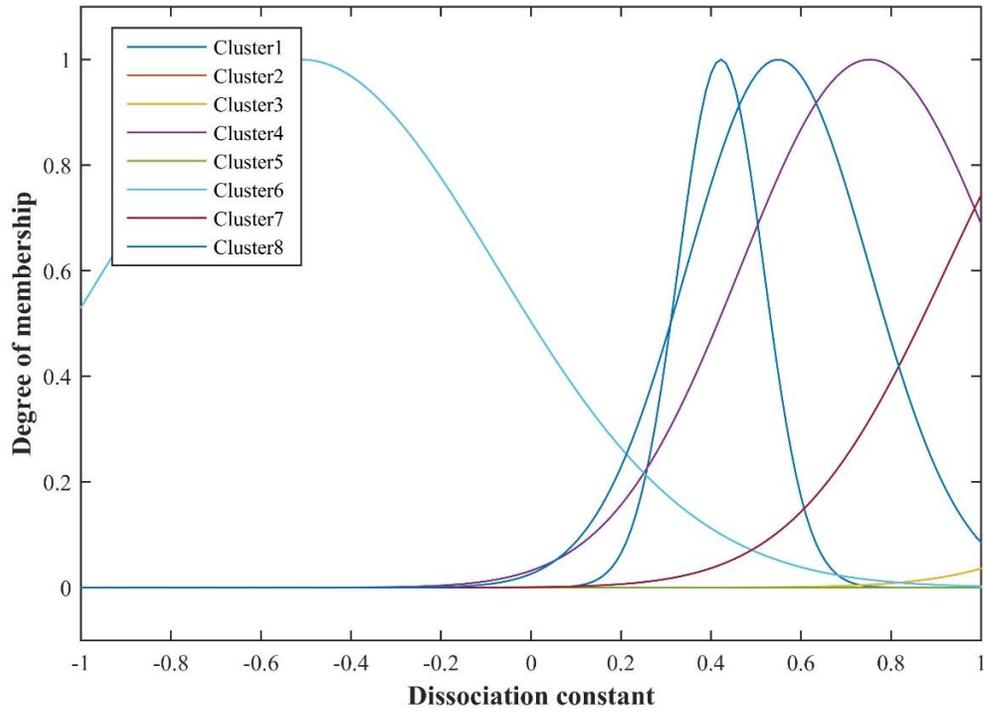

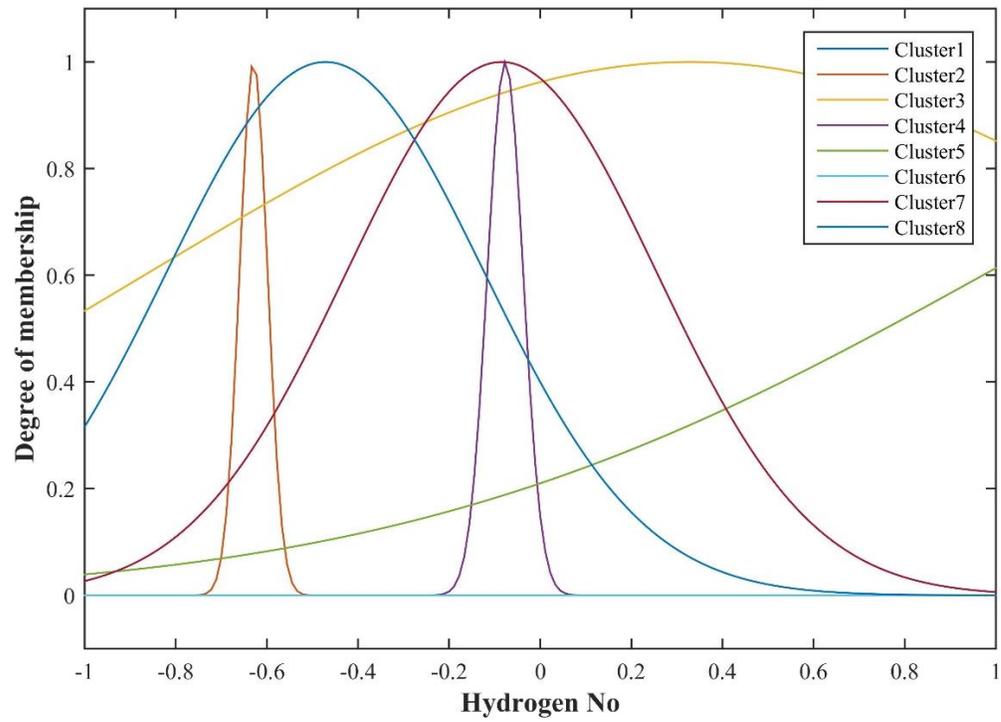

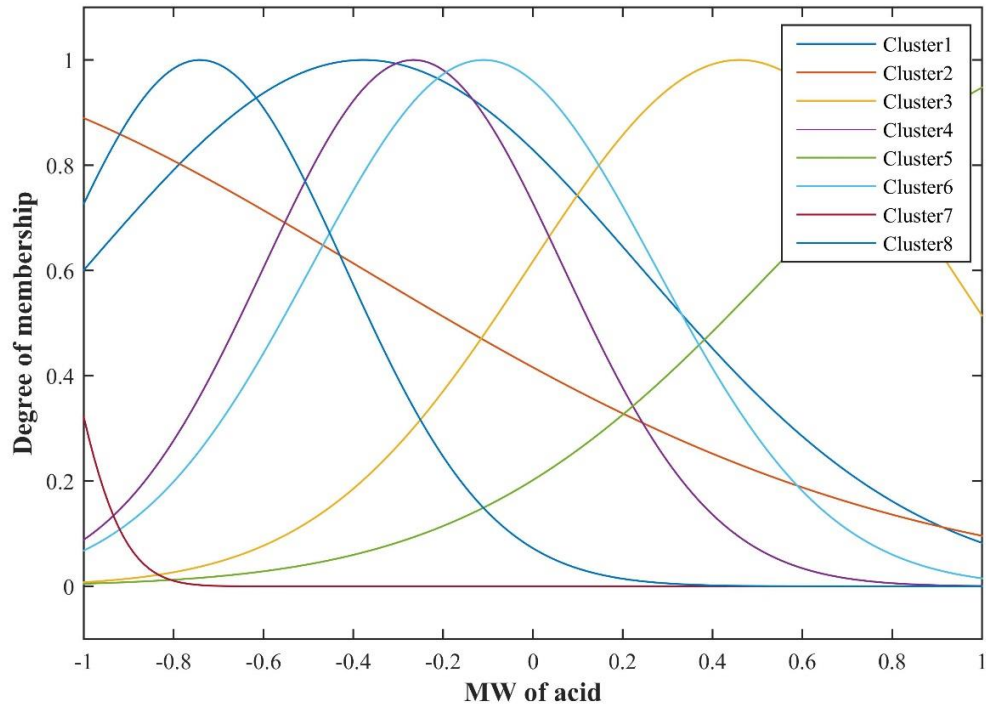

**Figure 5:** Trained membership function parameters

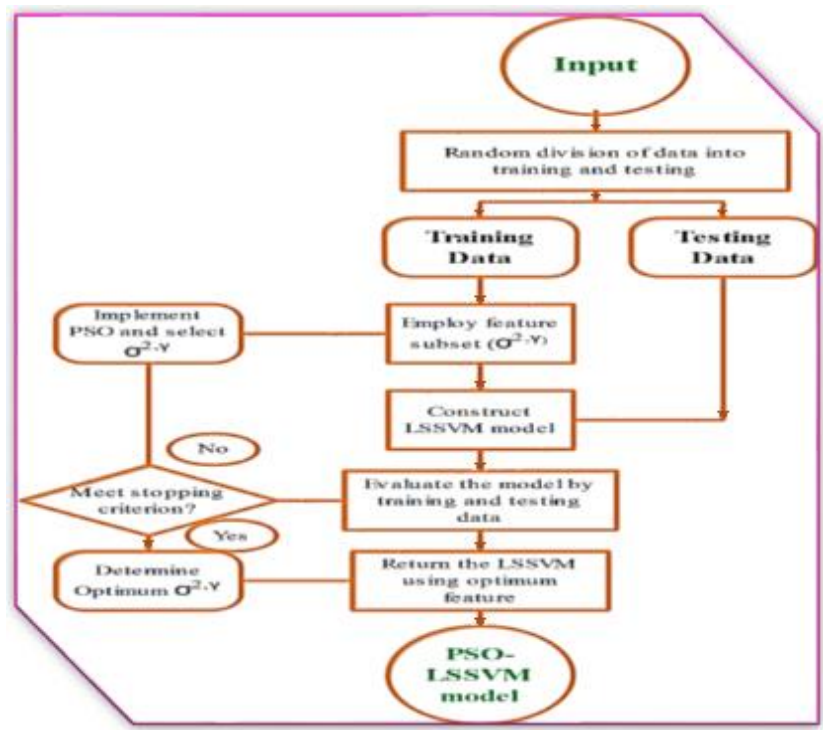

**Figure 6:** Schematic demonstration of trained LSSVM algorithm

The RBF kernel function due to its high degree of performance is utilized to construct the LSSVM algorithm. The LSSVM algorithm has two tuning parameters, $\sigma^2$ and $\Upsilon$ which are determined by utilizing PSO algorithm. The schematic demonstration of LSSVM algorithm is depicted in Fig. 6. The details of predicting models are summarized in Tab. 1. These details can be helpful in development of models for prediction of acid solubility in carbon dioxide.

**Table 1:** Details of proposed models

| Type | comment/value | Type | comment/value |
| --- | --- | --- | --- |
| **LSSVM** | | **ANFIS** | |
| **Kernel function** | RBF | Membership function | Gaussian |
| $\sigma^2$ | 0.80321 | No. of membership function parameters | 112 |
| $\Upsilon$ | 12893.2264 | No. of clusters | 8 |
| **Number of data utilized for training** | 141 | Number of data utilized for training | 141 |
| **Number of data utilized for testing** | 47 | Number of data utilized for testing | 47 |
| **Population size** | 85 | Population size | 50 |
| **Iteration** | 1000 | Iteration | 1000 |
| **C1** | 1 | C1 | 1 |
| **C2** | 2 | C2 | 2 |
| **MLP-ANN** | | **RBF-ANN** | |
| **No. input neuron layer** | 6 | No. input neuron layer | 6 |
| **No. hidden neuron layer** | 8 | No. hidden neuron layer | 50 |
| **No. output neuron layer** | 1 | No. output neuron layer | 1 |
| **Hidden layer activation function** | Sigmoid | Hidden layer activation function | RBF |
| **output layer activation function** | linear | output layer activation function | linear |
| **Number of data utilized for training** | 141 | Number of data utilized for training | 141 |
| **Number of data utilized for testing** | 47 | Number of data utilized for testing | 47 |
| **Number of max iteration** | 1500 | Number of max iteration | 50 |

In order to show the performance of proposed models in prediction of solubility of different acids, regression plots of RBF-ANN, MLP-ANN, ANFIS and LSSVM algorithms are depicted in Fig. 7 to compare the determined and actual solubility values. Based on these plots, the surprising fits for the predicting algorithms are obtained. Also, the predicted acid solubility data for proposed models are demonstrated along with the corresponding actual acid solubility values in Fig. S1. It can be observed that the model's output solubility values have excellent agreement with actual solubility values. Another graphical evaluation method is demonstration of relative error between predicted and experimental acid solubility in supercritical carbon dioxide. Fig. S2 shows the percentage of absolute error for the different predicting algorithm. The percentages of absolute error place under 1.5 percent for all developed algorithms, which expresses the acceptable degree of accuracy in prediction of acid solubility.

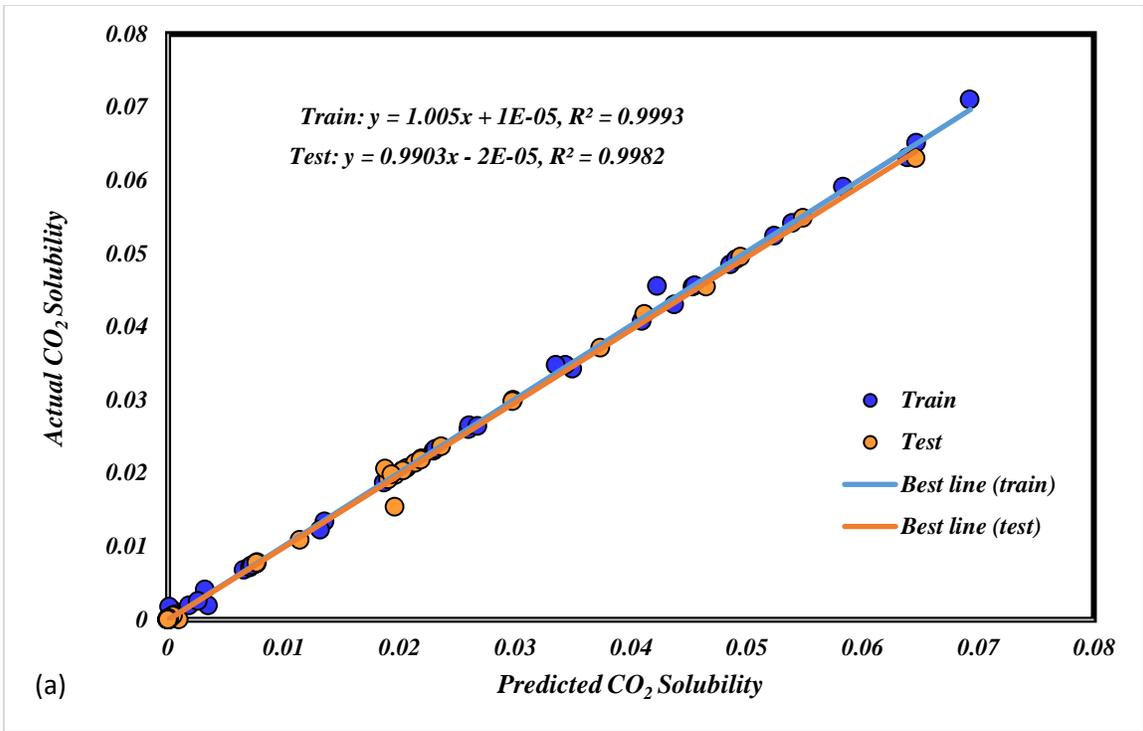

(a)

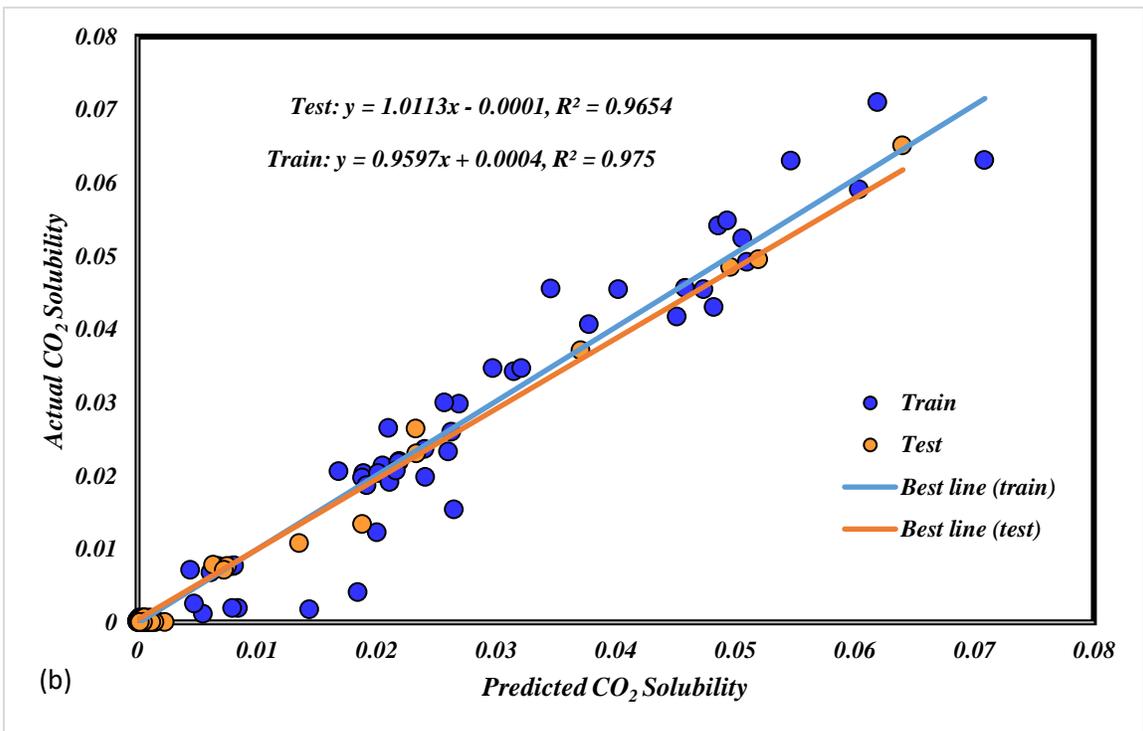

(b)

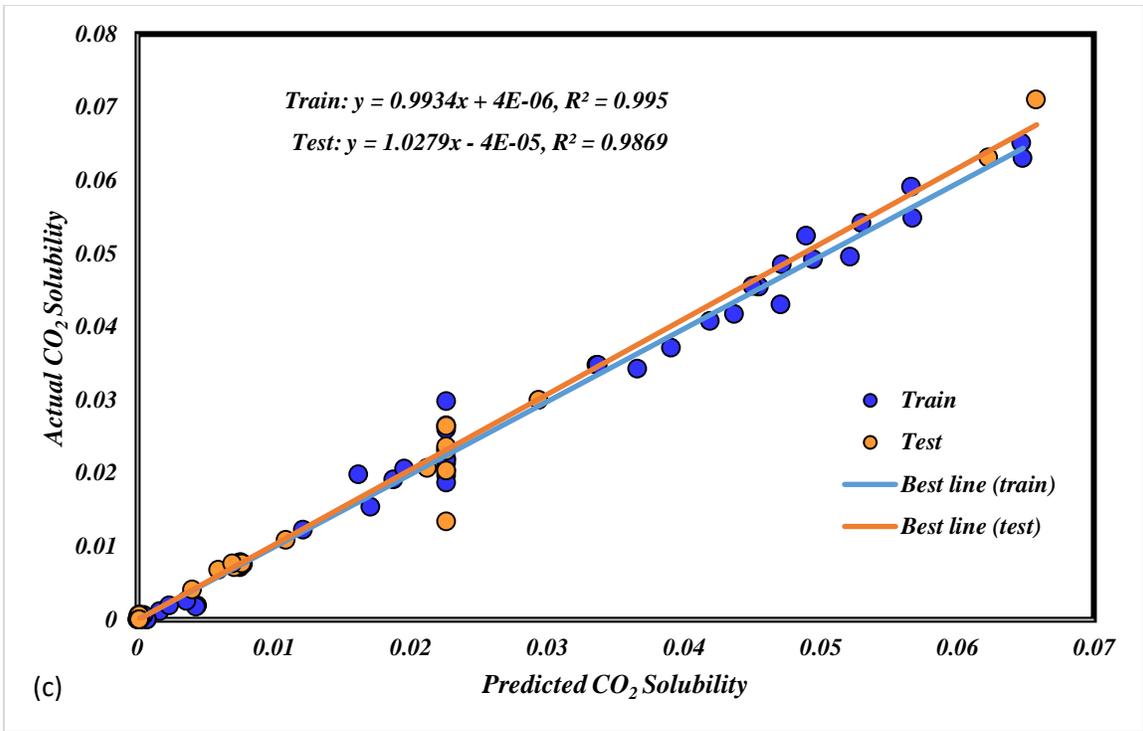

(c)

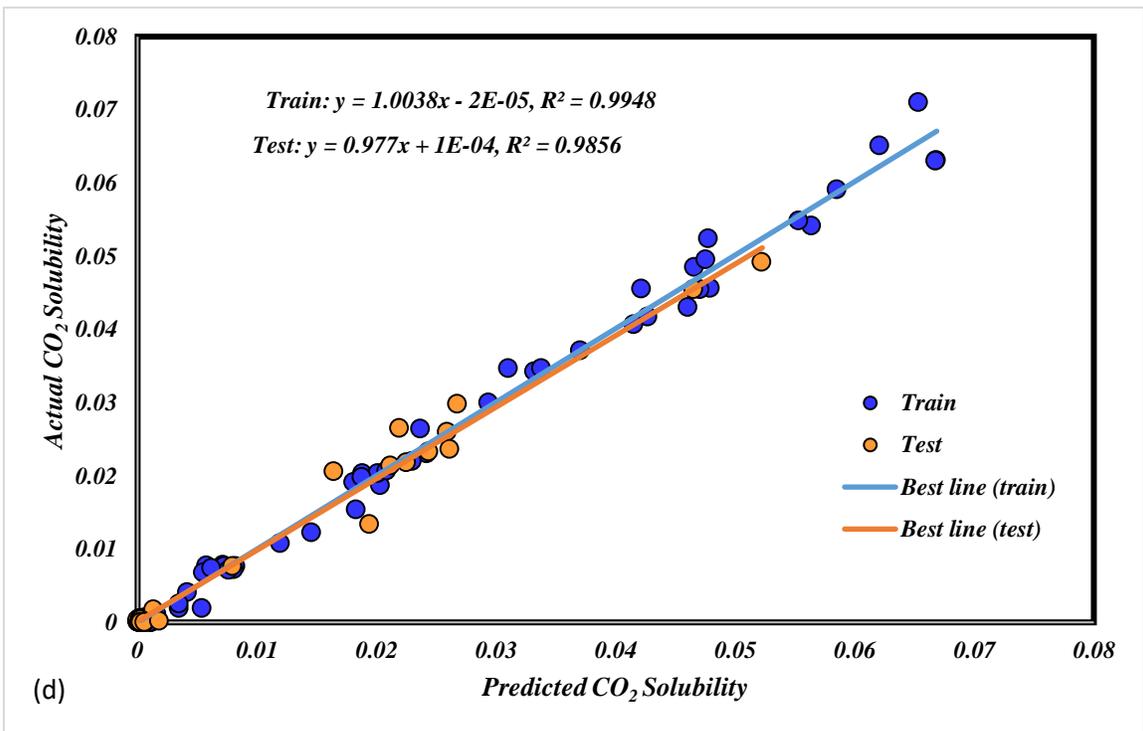

(d)

**Figure 7:** Regression plots obtained for different models

Furthermore, in order to clarify the performance of predicting algorithms, the statistical analysis is required so the coefficients of determination ($R^2$), average absolute deviation (AAD), Mean squared errors (MSEs) and Standard deviations (STDs) are determined such as following:

$$R^2 = 1 - \frac{\sum_{i=1}^{N}(X_i^{actual} - X_i^{predicted})^2}{\sum_{i=1}^{N}(X_i^{actual} - \overline{X^{actual}})^2} \quad \text{Eq. (18)}$$

$$AAD = \frac{1}{N}\sum_{i=1}^{N}|X_i^{predicted} - X_i^{actual}| \quad \text{Eq.(19)}$$

$$MSE = \frac{1}{N}\sum_{i=1}^{N}(X_i^{actual} - X_i^{predicted})^2 \quad \text{Eq. (20)}$$

$$STD_{error} = (\frac{1}{N-1}\sum_{i=1}^{N}(error - \overline{error}))^{0.5} \quad \text{Eq. (21)}$$

The $R^2$, AD, MSE and STD values of different algorithms are summarized in **Tab. 2**. According to these results, the LSSVM model has the greatest ability in forecasting acid solubility. The determined $R^2$ values for LSSVM is equal to 0.998 and 0.999 in train and test set, respectively. Furthermore it's RMSE, MSE and AAD parameters are 0.000527, 2.77875E-07, and 0.0179, respectively. According to these analyses LSSVM algorithm is known as the best predictor for prediction of solubility of different acids.

**Table 2:** Statistical analyses of models

| Model | Set | MSE | RMSE | $R^2$ | STD | AAD (%) |
| --- | --- | --- | --- | --- | --- | --- |
| **LSSVM** | Train | 5.72159E-07 | 0.000756 | 0.998 | 0.0007 | 0.0269 |
| | Test | 1.7978E-07 | 0.000424 | 0.999 | 0.0004 | 0.0149 |
| | Total | 2.77875E-07 | 0.000527 | 0.999 | 0.0005 | 0.0179 |
| **ANFIS** | Train | 5.79633E-06 | 0.002408 | 0.975 | 0.0022 | 0.1093 |
| | Test | 1.00976E-05 | 0.003178 | 0.965 | 0.0027 | 0.1677 |
| | Total | 9.02227E-06 | 0.003004 | 0.967 | 0.0026 | 0.1531 |
| **MLP-ANN** | Train | 3.23782E-06 | 0.001799 | 0.987 | 0.0017 | 0.0756 |
| | Test | 1.44839E-06 | 0.001203 | 0.995 | 0.0010 | 0.0600 |
| | Total | 1.89575E-06 | 0.001377 | 0.993 | 0.0012 | 0.0639 |
| **RBF-ANN** | Train | 2.33037E-06 | 0.001527 | 0.986 | 0.0013 | 0.0827 |
| | Test | 1.61993E-06 | 0.001273 | 0.995 | 0.0010 | 0.0779 |
| | Total | 1.79754E-06 | 0.001341 | 0.993 | 0.0011 | 0.0791 |

In addition to previous statistical indexes, there is another statistical approach to evaluate the reliability and accuracy of predicting algorithm, which called Leverage method. The mentioned approach consists of some statistical concepts such as model residuals, Hat matrix, and Williams plot which are used for detection of suspected and outlier data. There is more description of Leverage method in the literature (Rousseeuw and Leroy 2005).In this method, the residuals are estimated and inputs are utilized to build a matrix called Hat matrix such as follow:

$$H = X(X^T X)^{-1} X^T \qquad \text{Eq. (22)}$$

Where $X$ is the $m \times n$ matrix which n and m are the numbers of model parameters and samples respectively.

**Fig. 8** illustrates the William plot for the proposed models. As shown in this figure, the most of data points are in the range of leverage limit of residuals for -3 to 3. The leverage limit is formulated such as following:

$$H^* = 3(n + 1)/m \qquad \text{Eq. (23)}$$

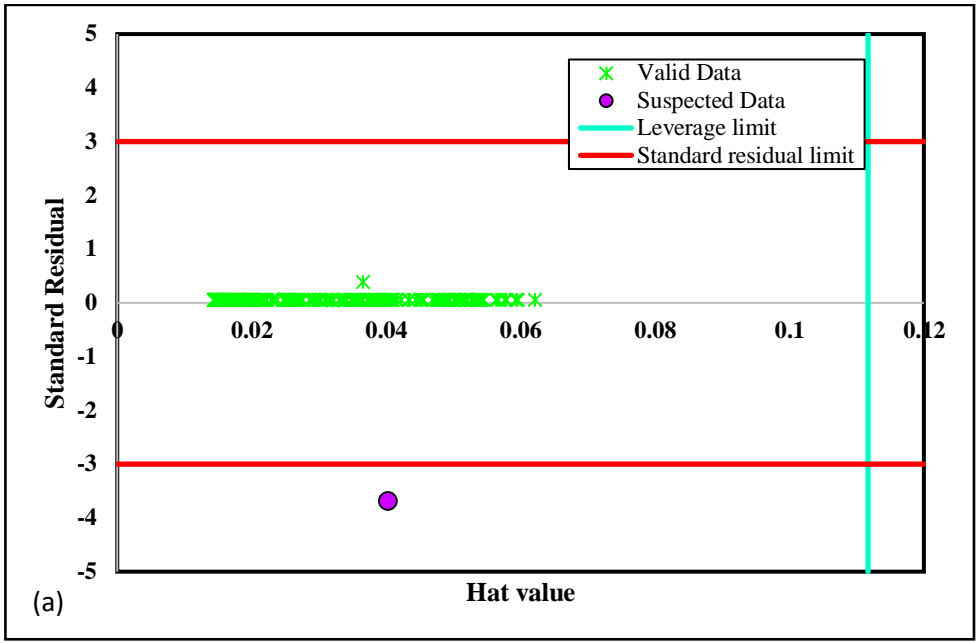
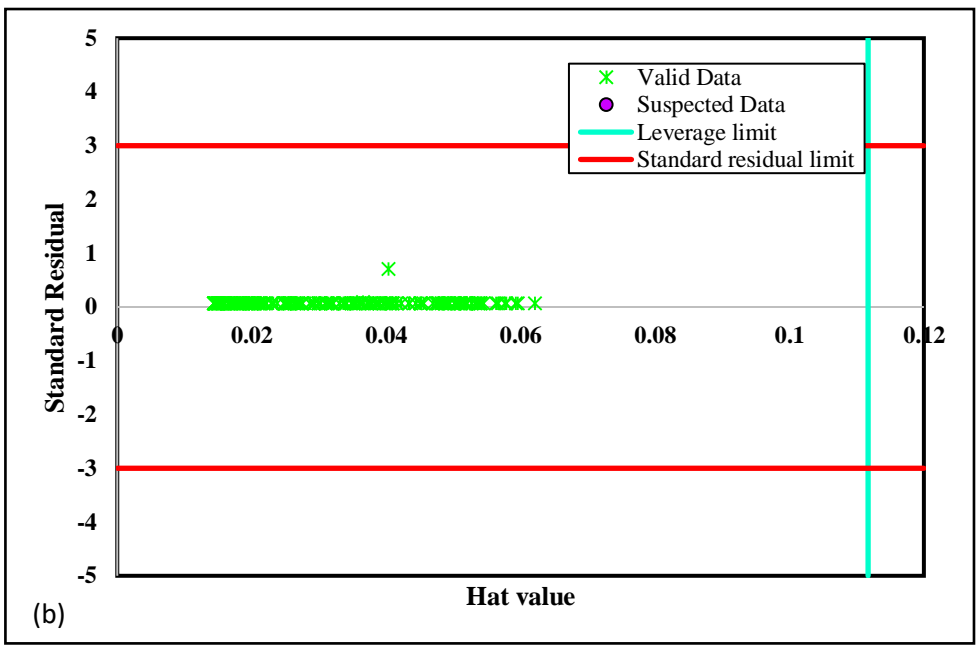

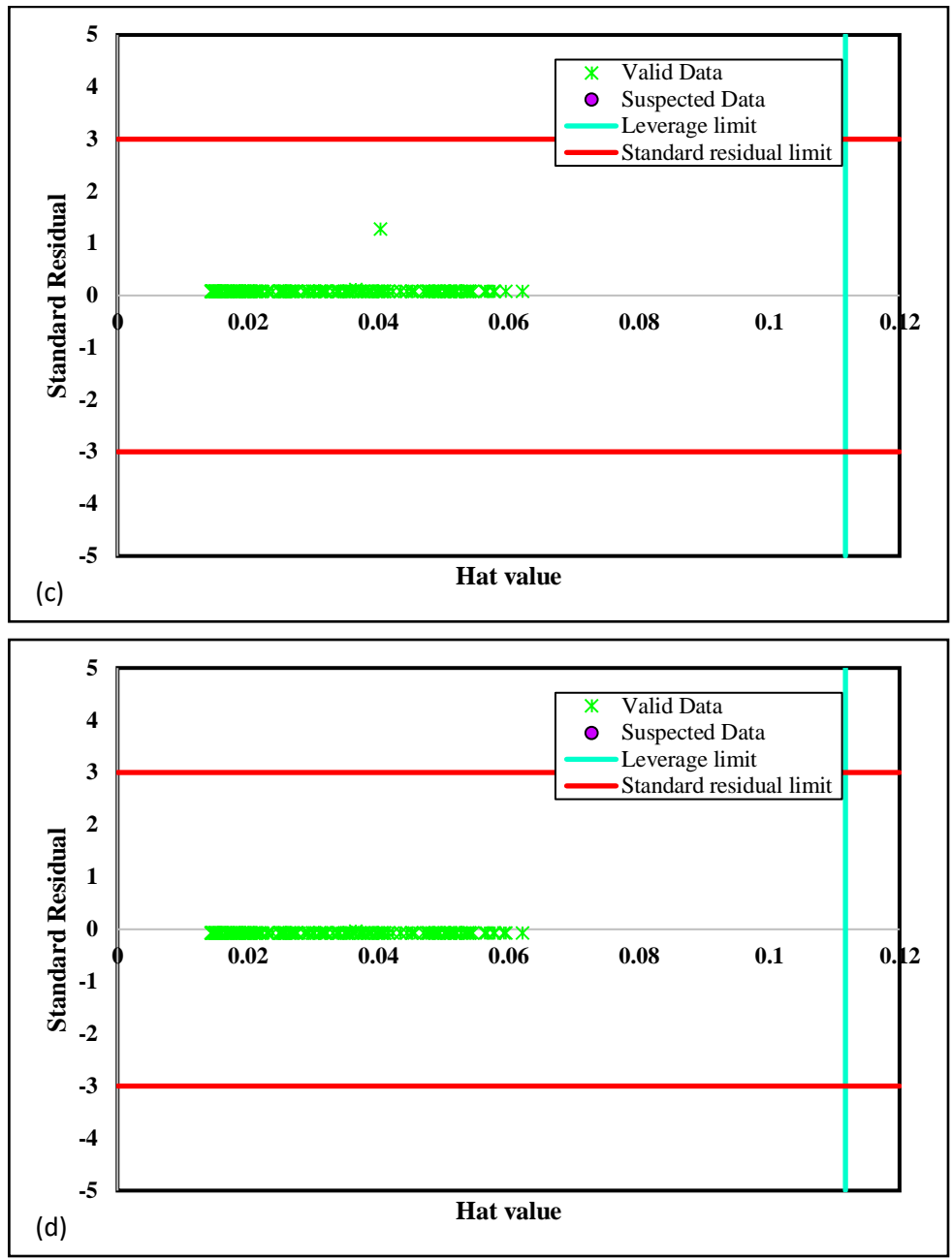

**Figure 8:** Absolute deviation plots for (a) LSSVM, (b) ANFIS, (c) MLP-ANN, and (d) RBF-ANN

Another method to investigate the validity of the models is a parametric analysis of solubility. To this end, the Relevancy index is introduced to investigate the impact of inputs on acid solubility. The Relevancy index is determined such as following (Zarei, Razavi et al. 2019):

$$r = \frac{\sum_{i=1}^{n}(X_{k,i}-\overline{X_k})(Y_i-\overline{Y})}{\sqrt{\sum_{i=1}^{n}(X_{k,i}-\overline{X_k})^2 \sum_{i=1}^{n}(Y_i-\overline{Y})^2}}$$

Eq. (24)

where $Y_i$, $\overline{Y}$, $X_{k,i}$ and $\overline{X_k}$ are the 'i' th output, output average, kth of input and average of input. The Relevancy index absolute value represent the effectiveness of the parameters on acid solubility. As shown in **Fig. 9**, the molecular weight of acid has the most Relevancy factor between different input parameters so this parameter is known as the most effective parameters on acid solubility in supercritical carbon dioxide. Moreover, acid dissociation constant has the least effect on acid solubility. This figure illustrates that as number of carbon and hydrogen of acid, molecular weight and pressure increase, acid solubility in carbon dioxide increases. On the other hand, increasing acid dissociation constant and temperature caused drop in solubility of acid in carbon dioxide.

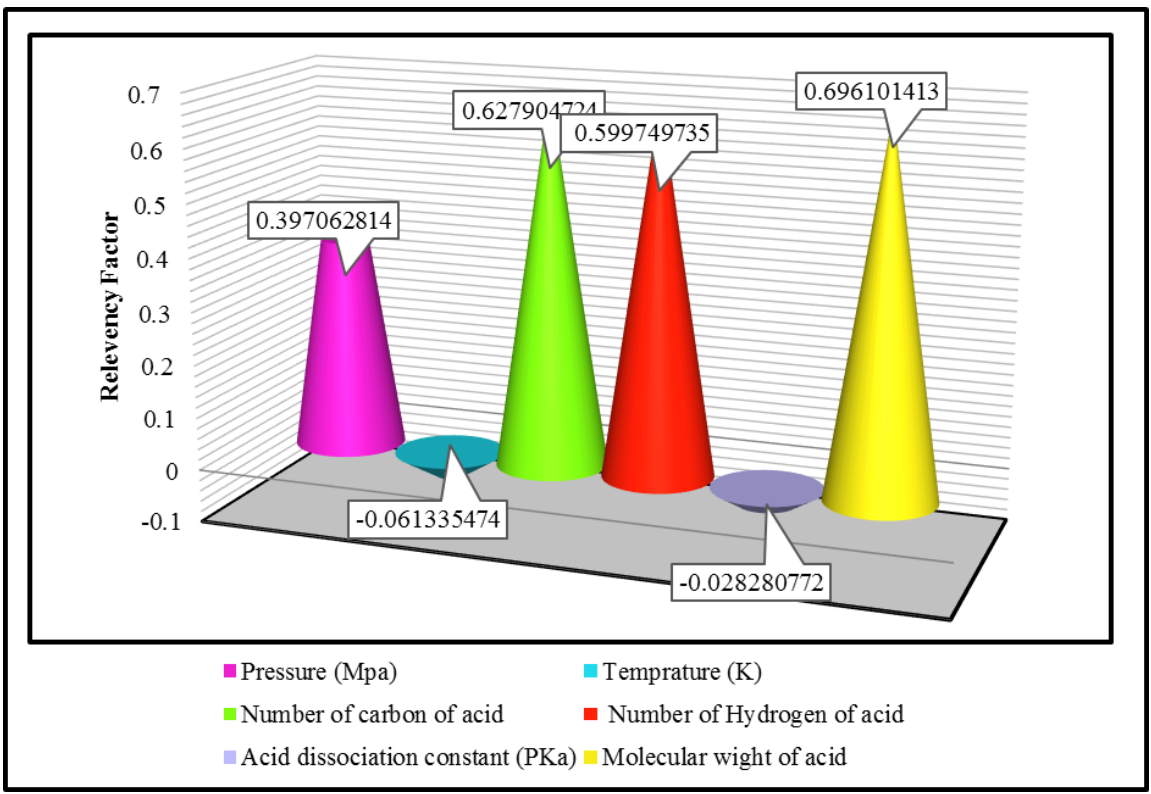

**Figure 9:** Sensitivity analysis of investigated variables

## 4  Conclusions

In this paper, we have applied RBF-ANN, MLP-ANN, ANFIS-PSO and LSSVM algorithms to determine the different acids solubility values in supercritical carbon dioxide in terms of pressure, temperature, and different acid structure based on a reliable databank which gathered from the literature. These predicting approaches can forecast acid solubility in the wide range of operating conditions. To prove the aforementioned acclaim, different statistical and graphical evaluations have been performed in the previous section. According to the obtained results from comparisons, the LSSVM model has the best performance respect to the others and ANFIS algorithm has the least of accuracy in this prediction. Also, the results of sensitivity analysis identify the molecular weight of the acid parameter is the most effective factor in solubility of acids in supercritical carbon dioxide. Based on these comprehensive investigations this manuscript has great potential and ability to help the researchers in their future works.

**Nomenclature**

| | |
|---|---|
| **ANFIS** | Adaptive neuro-fuzzy inference system |
| **LSSVM** | Least squares support vector machine |
| **RBF-ANN** | Radial basis function artificial neural network |
| **MLP-ANN** | Multi-layer Perceptron artificial neural network |
| **PSO** | Particle swarm optimization |
| **φ(x)** | nonlinear function |
| **ω** | weight |
| **b** | bias |
| **ϒ** | regularization parameter |
| **$e_k$** | support value |
| **K** | kernel function |
| **Z** | Gaussian parameter |
| **σ** | Gaussian parameter |
| **m** | One of the resulting index of ANFIS |
| **n** | One of the resulting index of ANFIS |
| **r** | One of the resulting index of ANFIS |
| **W** | inertia weight |
| **c** | learning rate |
| **$R^2$** | coefficient of determination |
| **AAD** | average absolute deviation |
| **MSE** | Mean squared error |
| **STD** | Standard deviation |
| **H** | Hat matrix |
| **$H^*$** | The leverage limit |

## Supplementary content

**Table S1.** Experimental data which are used in this study

| Acid name | Pressure | Temperature (K) | Acid dissociation constant (PKa) | solubility (mol/mol) | No of data points | References |
|---|---|---|---|---|---|---|
| **Perfluoropentanoic acid** | 10-26.2 | 314-334 | 0.52 | 0.0134-0.0298 | 17 | (Dartiguelongue, Leybros et al. 2016) |
| **o-Hydroxybenzoic Acid** | 8.11-20.26 | 308.15-328.15 | 4.06 | 0.000007-0.000624 | 49 | (Gurdial and Foster 1991) |
| **Corosolic Acid** | 8.0-30 | 308.15-333.15 | 4.7 | $3.28*10^{-11}$ - 0.071 | 40 | (Kumoro 2011) |
| **Maleic Acid** | 7.0-30 | 318.15-348.15 | 1.83 | 0.000013-0.0005917 | 21 | (Sahihi, Ghaziaskar et al. 2010) |
| **Ferulic Acid** | 12.0-28 | 301.15-333.15 | 4.38 | 0.00000155-0.0000118 | 18 | (Sovova, Zarevucka et al. 2001) |
| **Azelaic Acid** | 10.0-30 | 313.15-333.15 | 4.84 | 0.00000042-0.00001012 | 14 | (Sparks, Hernandez et al. 2007) |
| **Nonanoic Acid** | 10.0-30 | 313.15-333.15 | 4.96 | 0.00013-0.00782 | 14 | (Sparks, Estévez et al. 2008) |
| **p-aminobanzoic acid** | 8.0-21 | 308-328.0 | 4.78 | 0.000001302-0.000006452 | 15 | (Tian, Jin et al. 2007) |
| | | | | | Total= 188 | |

**Table S2.** Average of experimental data which are used in this study

| Acid name | Pressure (Mpa) | Temprature (K) | solublity(mol/mol) |
|---|---|---|---|
| Perfluoropentanoic acid | 17.37058824 | 324 | 0.022118 |
| o-Hydroxybenzoic Acid | 13.84040816 | 316.6193878 | 0.000238 |
| Corosolic Acid | 18.2 | 319.4 | 0.029932 |
| Maleic Acid | 16.42857143 | 333.15 | 0.000173 |
| Ferulic Acid | 19.83333333 | 319.4833333 | 5.37E-06 |
| Azelaic Acid | 20 | 323.15 | 3.92E-06 |
| Nonanoic (Pelargonic) Acid | 20 | 323.15 | 0.006548 |
| p-aminobanzoic acid | 14 | 318 | 3.82E-06 |

**Table S3.** Details of acids which are utilized in this investigation.

| Acid name | structure | Empirical Formula or linear formula | Molecular weight gr/mole |
|---|---|---|---|
| Perfluoropentanoic acid | 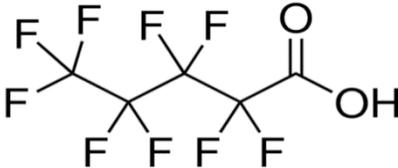 | CF$_3$(CF$_2$)$_3$COOH | 264.05 |
| o-Hydroxybenzoic Acid | 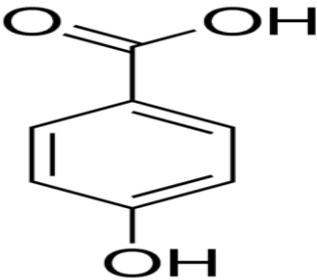 | HOC$_6$H$_4$CO$_2$H | 138.12 |
| Corosolic Acid | 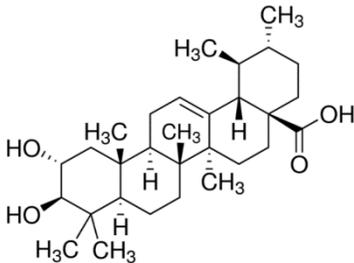 | C$_{30}$H$_{48}$O$_4$ | 472.70 |
| Maleic Acid | 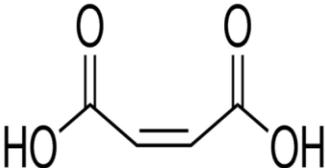 | HO$_2$CCH=CHCO$_2$H | 116.07 |
| Ferulic Acid | 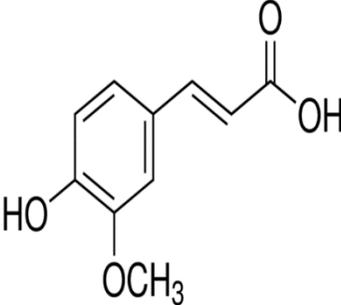 | HOC$_6$H$_3$(OCH$_3$)CH=CHCO$_2$H | 194.18 |
| Azelaic Acid | 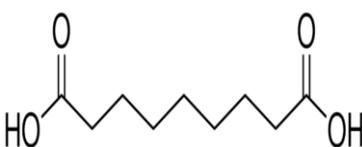 | HO$_2$C(CH$_2$)$_7$CO$_2$H | 188.22 |

| Nonanoic (Sparks, Estévez et al. 2008) Acid | 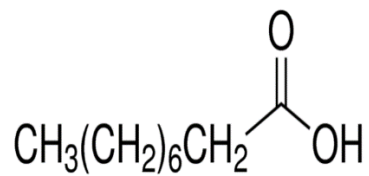 | CH$_3$(CH$_2$)$_7$COOH | 158.24 |
| --- | --- | --- | --- |
| p-aminobanzoic acid | 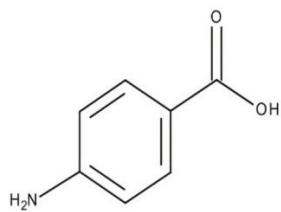 | C$_7$H$_7$NO$_2$ | 137.14 |

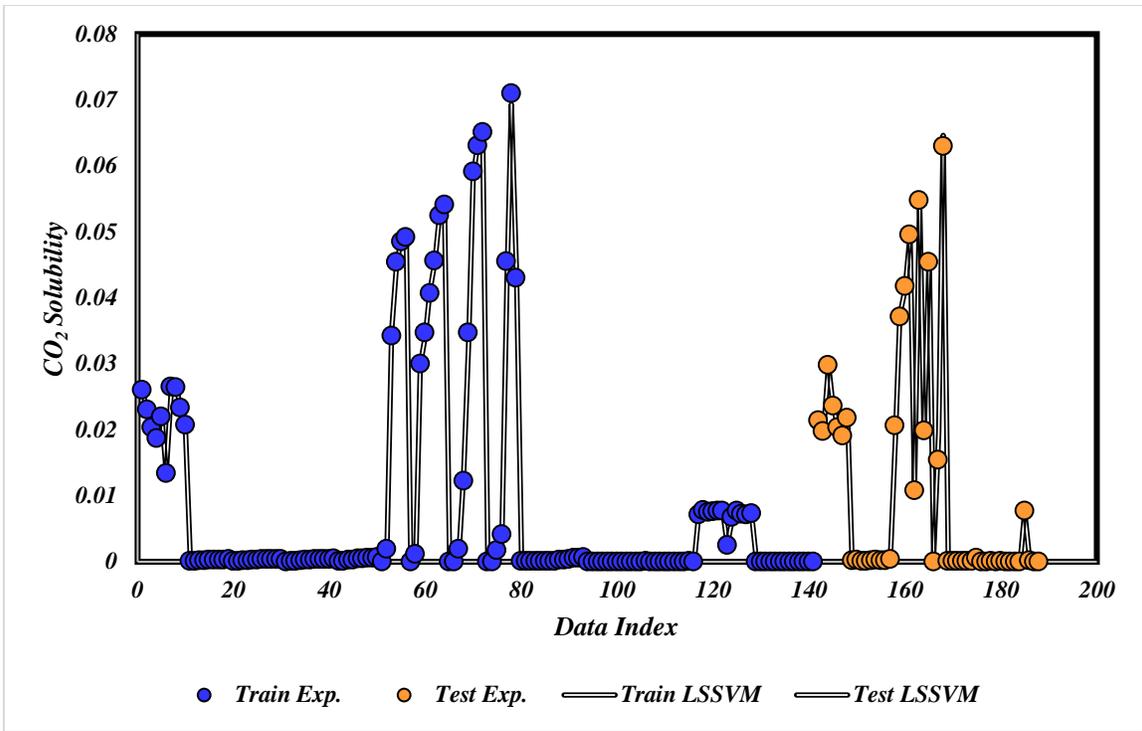
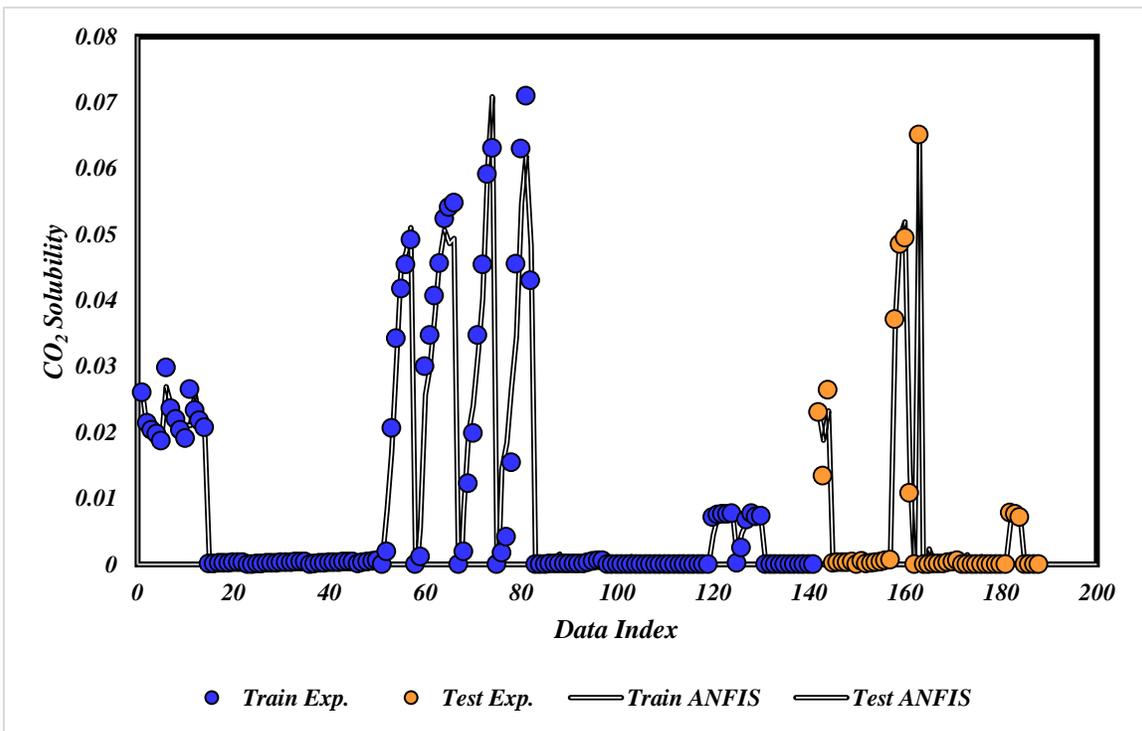

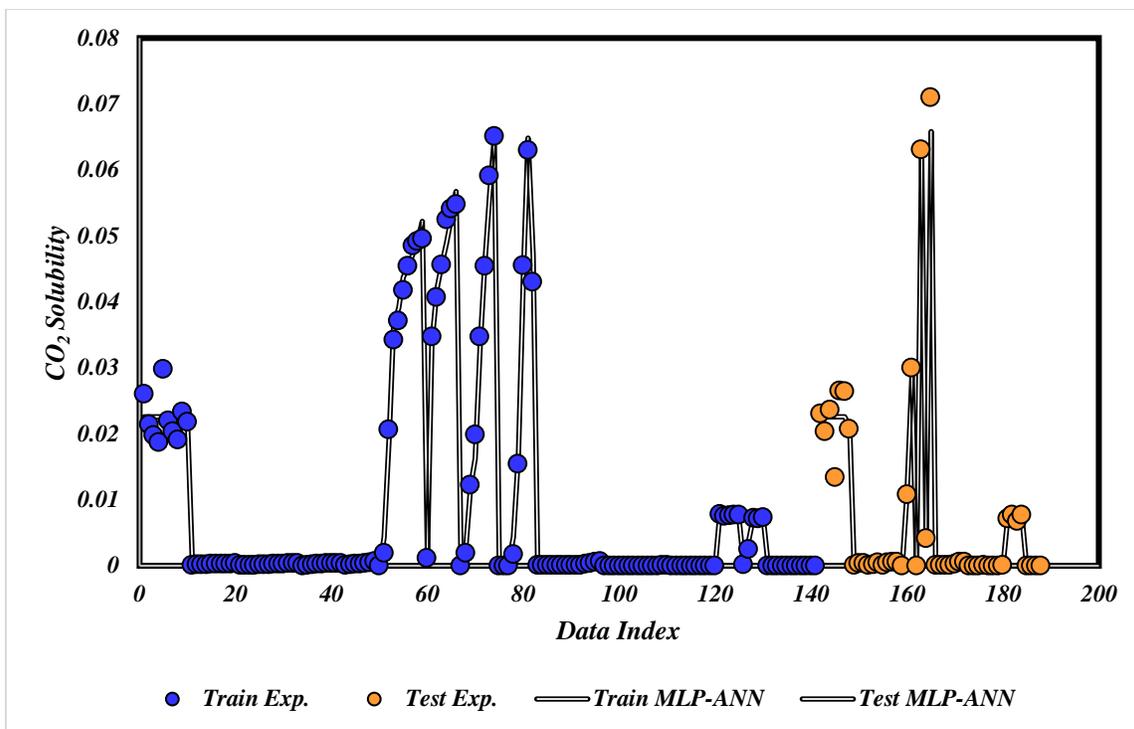
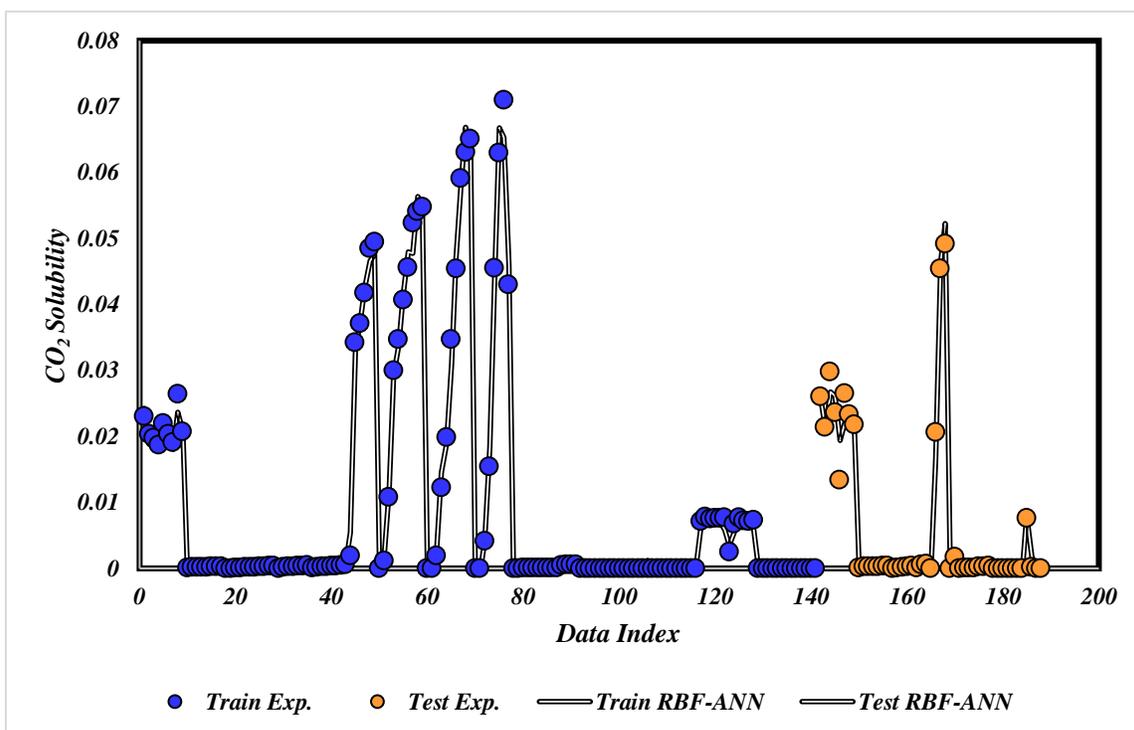

**Figure S1:** Experimental and predicted solubility of $CO_2$ by the proposed models

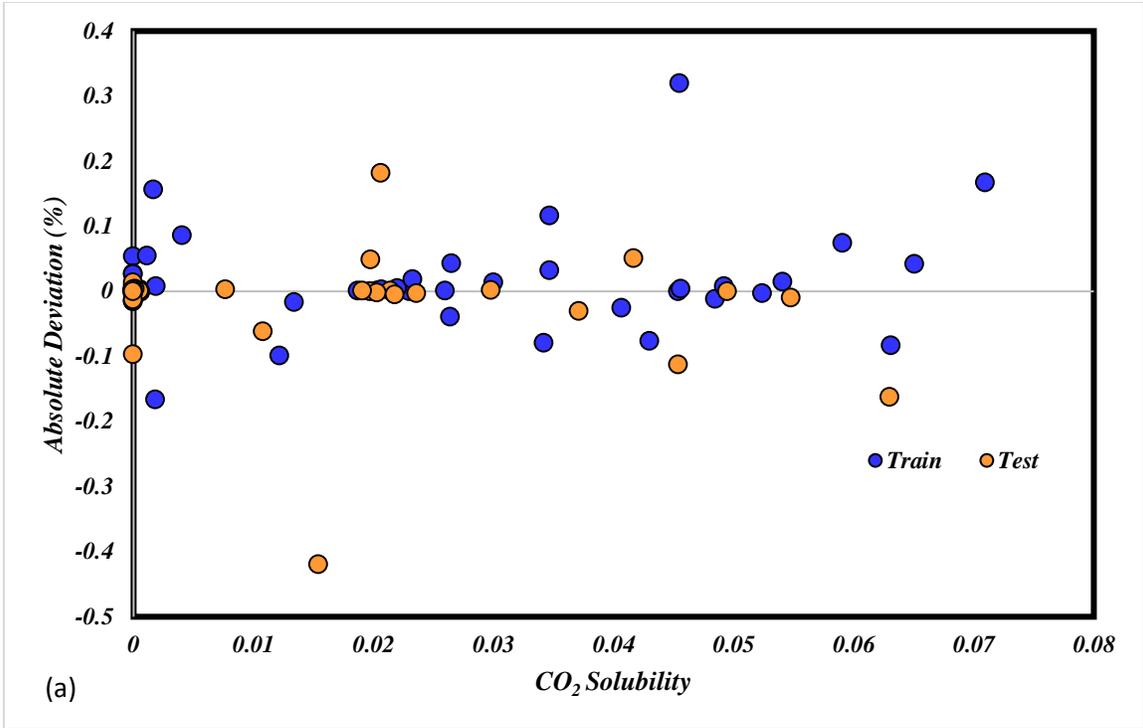

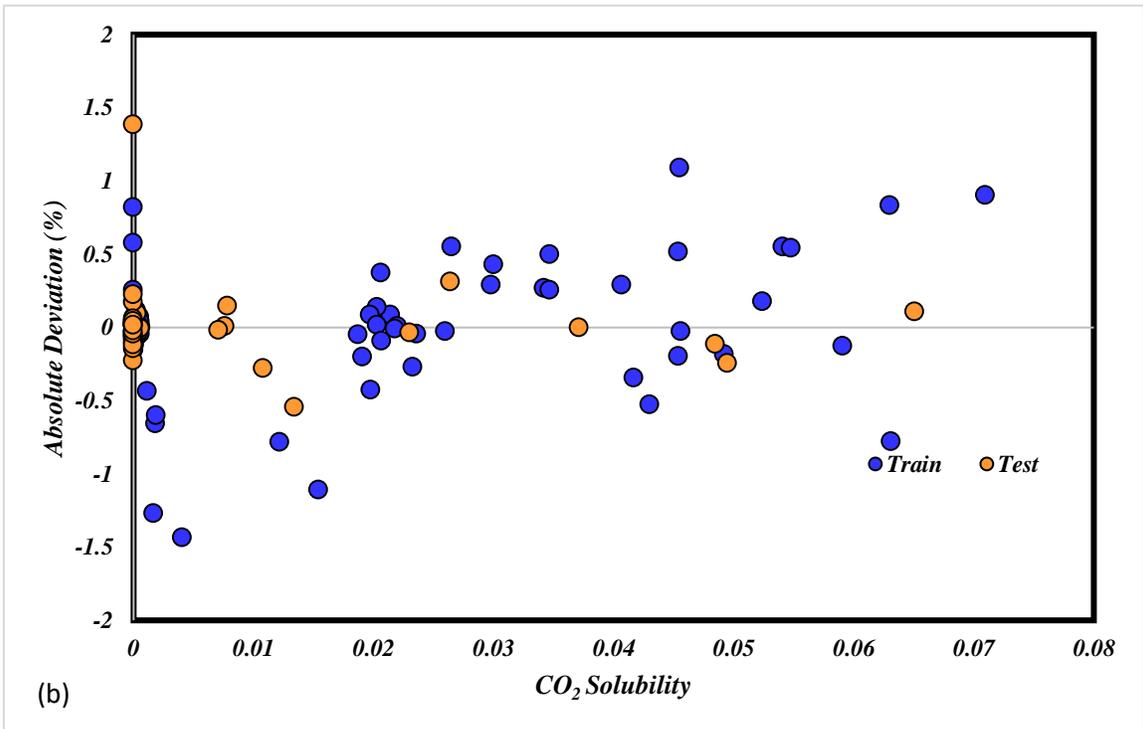

(a)

(b)

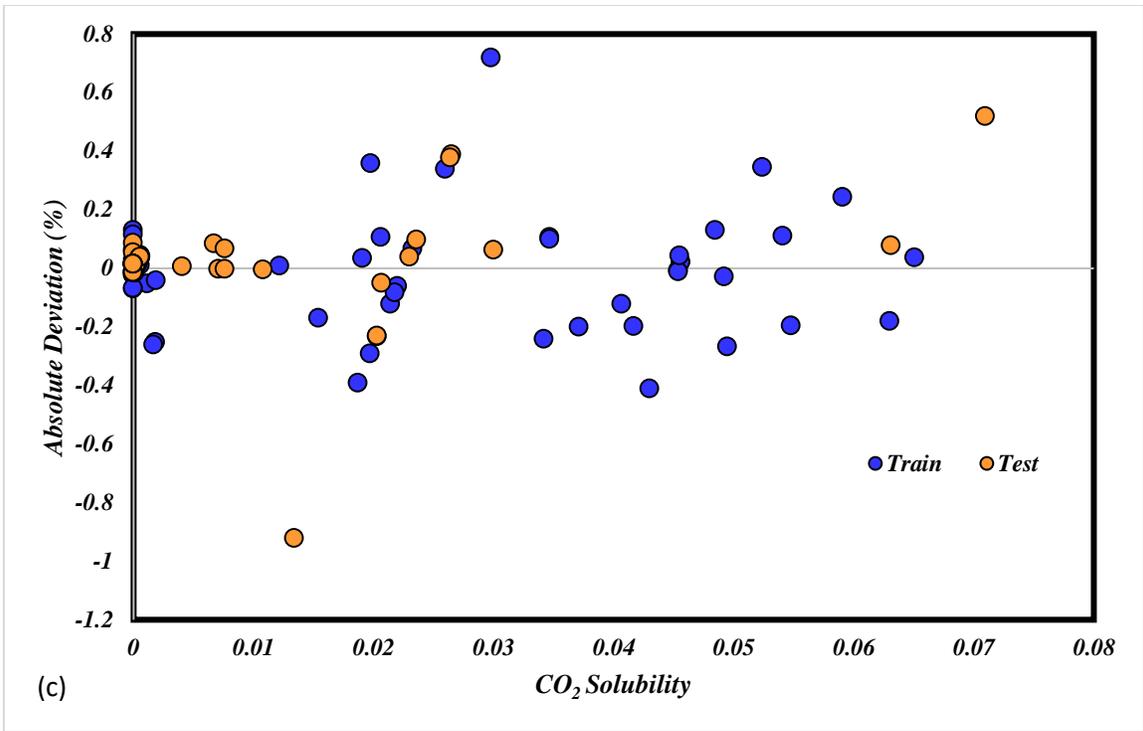
(c)

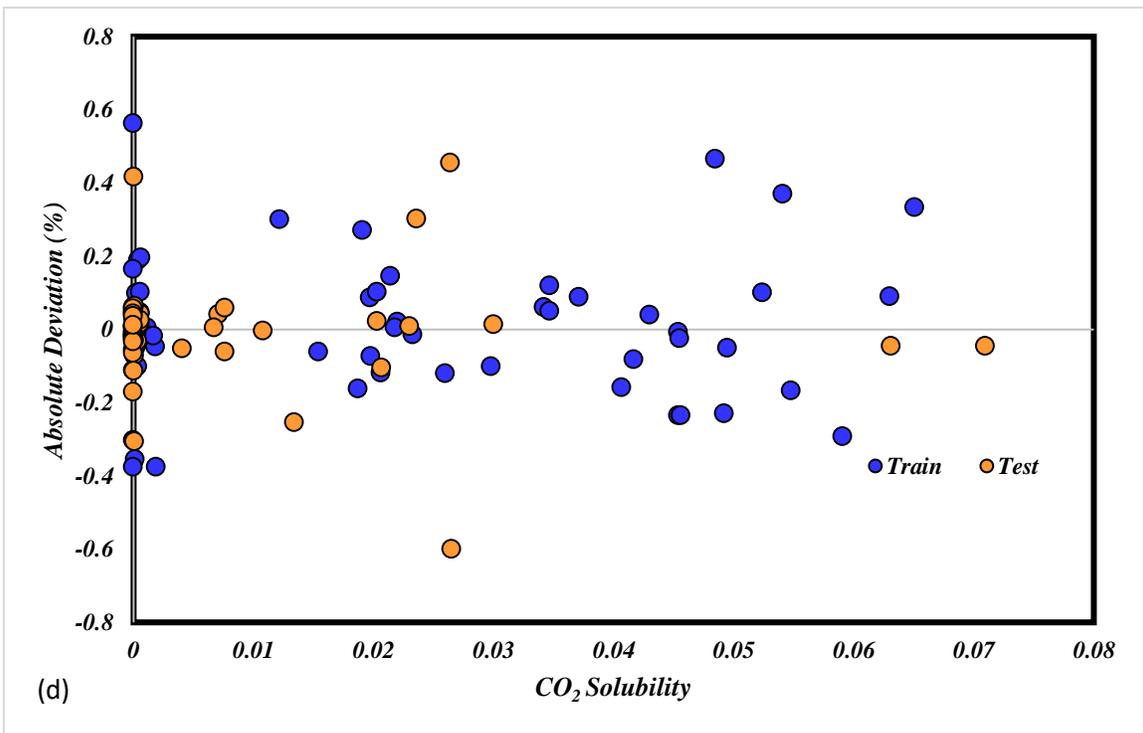
(d)

**Figure S2:** Absolut deviation plots for (a) LSSVM, (b) ANFIS, (c) MLP-ANN, and (d) RBF-ANN

**Acknowledgement**

This publication has been supported by the Project: "Support of research and development activities of the J. Selye University in the field of Digital Slovakia and creative industry" of the Research & Innovation Operational Programme (ITMS code: NFP313010T504) co-funded by the European Regional Development Fund.